\documentclass[letter]{aa} 
\usepackage{graphicx}
\usepackage{txfonts}
\begin{document}
\title{The young, wide and very low mass visual binary LOri167\thanks{Based on observations 
 carried out with the CFHT, the WHT, the CAHA, 
the Spitzer and the Keck telescopes}
}

   \author{D. Barrado y Navascu\'es
          \inst{1}
           \and
           A. Bayo
          \inst{1}
           \and
	   M. Morales-Calder\'on
          \inst{1}
           \and
	   N. Hu\'elamo
          \inst{1}
           \and
	   J.R. Stauffer
          \inst{2}
           \and
	   H. Bouy
          \inst{3}
          }

   \offprints{D. Barrado y Navascu\'es}

   \institute{Laboratorio de Astrofisica Espacial y Fisica Fundamental, LAEFF-INTA,
    Madrid, Spain\\
              \email{barrado@laeff.inta.es}
         \and
             IPAC, Caltech, Pasadena, USA\\
             \email{stauffer@ipac.caltech.edu}
         \and
             Astronomy Department, University of California,
               Berkeley, USA\\
             \email{hbouy@astro.berkeley.edu}
             }

   \date{2007;2007}


\abstract
%
{}
{We look for wide, faint companions around members of the 5 Myr Lambda Orionis open cluster.}
{We used optical, near-infrared,  and Spitzer/IRAC photometry.}
{We report the discovery of a very wide very low mass visual binary, LOri167, formed by a brown dwarf 
and a planetary-mass candidate located at 5 arcsec,
 which seems to belong to the cluster. We  derive T$_{eff}$ of
2125 and 1750 K. If they are members, comparisons with theoretical models indicate masses
of 17$^{+3}_{-2}$ and 8$^{+5}_{-1}$ $M_{jup}$, with a projected separation of 2000 AU. }
{Such a binary system would be difficult to explain in most models, particularly those
where substellar objects form in the disks surrounding higher mass stars.}
{}
\keywords{open clusters: individual (Collinder 69) -- Stars: low-mass, brown dwarfs}


   \maketitle
%

\section{Introduction}

During the past few months, several very low mass  binaries
have been reported in the literature 
(Close et al. 2003; Chauvin et al. 2004; 
Kraus et al. 2005, 2006). 
They are relatively close visual binaries, with angular separations 
of a fraction of an arcsecond and projected distances of a few tens of AU or less.
On the other hand, other authors 
(Bill{\`e}res et al. 2005; Luhman et al. 2004, 2005;
Jayawardhana \& Ivanov 2006; Close et al. 2006;
 Bouy et al. 2006; Caballero et al. 2007; Caballero 2007)
have found a population of very low mass  binaries in several 
young associations or in the field, with separations in the range 100-250 AU.
These binaries might include either one or two brown dwarfs, or might also
include a planetary-mass object, whose mass is predicted by theoretical models to be below the
deuterium--burning limit at about 13 $M_{jup}$ masses. Such binaries are extremely important
because they challenge the latest formation scenarios for very low mass objects 
and might indicate that very low mass brown dwarfs and planetary-mass objects form
by collapse and fragmentation of molecular clouds, just as stars do, and that the minimum
mass for the process is much lower than what had been previously thought.

We have discovered a close (5\arcsec) optical binary in the
Lambda Orionis cluster (Collinder 69). This association is about 
5 Myr (Dolan \&  Mathieu 1999, 2001; Barrado y Navascu\'es et al. 2007)
and is located at about 400 pc. If the pair is physically associated both
 with each other and with the cluster,
they would be one of the lowest mass binary systems and would have
one of the widest separations reported in the literature, about 2000 AU.

\section{Analysis}


We combined optical, near infrared,
 and Spitzer/IRAC photometry in order to look for faint,
 very low mass members of the Collinder 69 or Lambda Orionis open cluster (Dolan \& Mathieu 1999, 2001).
 We first combined optical data in 
the $Ic$ filter taken at the CFHT in October 1999 with the 12K camera (Barrado y Navascu\'es et al. 2004) 
and with deep $J$ imaging taken at the WHT in February 2003 with the INGRID instrument 
(Barrado y Navascu\'es, Hu\'elamo \& Morales-Calder\'on 2005). The use of a color-color diagram allowed us
to select a sample of very faint, red objects. These results will be reported elsewhere. 
In particular, one of them was very close to the brown dwarf candidate LOri167 
(Fig. 1, see also Fig. 1 in Barrado y Navascu\'es, Hu\'elamo \& Morales-Calder\'on 2005),
 whose mass, if it is a member, is
just above the planetary-mass borderline at 13 $M_{jup}$, in the range 
%
%
15-20 $M_{jup}$ --Dusty 5 Myr isochrone by 
Chabrier et al. (2000), which is appropriate for its effective temperature (see below).
Subsequent follow-up with the IRAC camera onboard the Spitzer
infrared telescope has helped us to confirm that
this object falls in the cluster sequence (see Fig. 2).

\begin{figure}[t]
\includegraphics[angle=0,scale=.40]{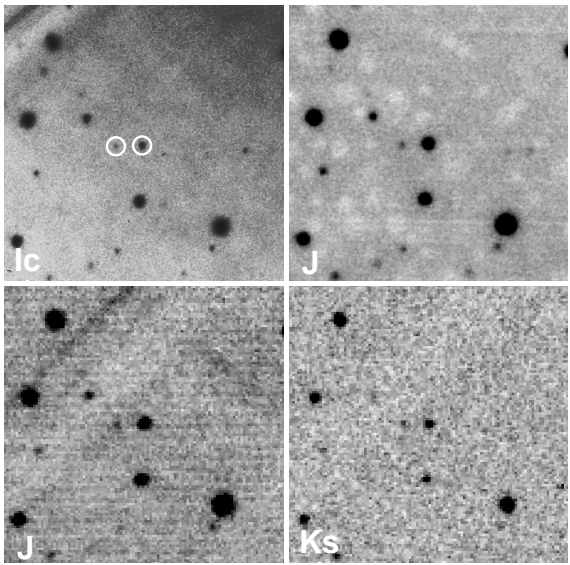}
\caption{45\arcsec$\times$45\arcsec \, images around LOri167, corresponding to the 
CFHT in october 1999  and the WHT in february 2003.}
\end{figure}

We also observed this area in the filters $J$, $H$, and $Ks$ in previous campaigns with WHT/INGRID
(November 2002) and the Calar Alto 3.5m telescope and Omega 2000 (October 2005).
 All the photometric data were analyzed in the same way, using aperture photometry within
the IRAF\footnote{IRAF is distributed by National Optical Astronomy Observatories,
 which is operated by the Association of Universities for Research in
 Astronomy, Inc., under contract to the National Science Foundation, USA} environment. 
For the IRAC data, we used an aperture of 2 pixels, 
to reduce the errors of the faint companion.
For the near-IR data,
the calibration was achieved using either standard stars or the 
2MASS catalog and the stars present in the field of view. The data are listed in Table 1. 
%

\subsection{Membership in the cluster}

Using all available data (see Table 1),
we were able to build the spectral energy distribution (SED) of both objects (Fig. 3). 
Clearly, they are cool objects, with the visual companion much cooler 
than LOri167. 
%
%
We tried to derive proper motions in our images, but the accuracy of 
the astrometry goes with the inverse of the signal-to-noise ratio, and the data do not
go deep enough for these two objects.


\begin{figure}[t]
\includegraphics[angle=0,scale=.40]{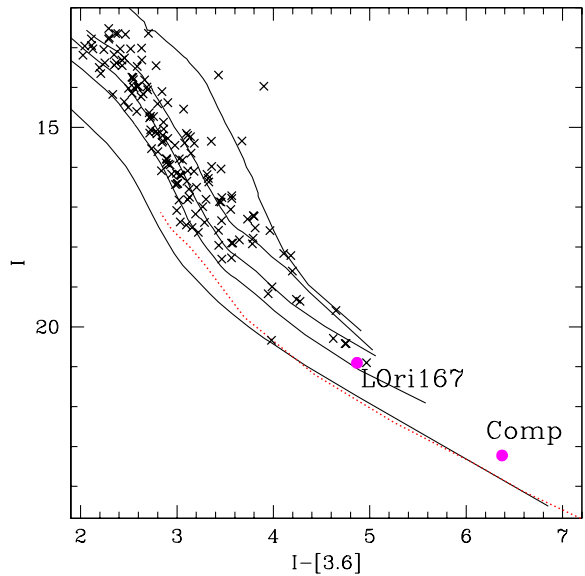}
\caption{ Color magnitude diagram  for the 5 Myr cluster  Collinder 69. 
The LOri167 brown dwarf and its visual companion  appear as thick, big  circles.
NextGen isochrones of 1, 5, 10, 20 and 100 Myr by Baraffe et al. (1998) appear as solid lines
($L$ data instead [3.6]).
We have also included a Dusty 5 Myr isochrone --dotted line.
}
\end{figure}

Despite the faintness of LOri167, we tried to
 secure a low-resolution spectrum  on a couple
of occasions, both with the Keck telescope and the
NIRSPEC spectrograph, in order to derive
a spectral classification (November 2004 and December 2005). 
Unfortunately 
on both occasiones the weather did not cooperate 
(bad seeing, clouds),
and we were only able to obtain a noisy, low-quality spectrum. 
Comparison with spectral templates suggests an M9-L2 spectral type for LOri167.
If it is a member of Collinder 69 (very likely), LOri167 would have, from comparison with 
theoretical isochrones, a mass in the range 15-20 $M_{jup}$, as stated before.
These estimates also include a maximum age of 7-8 Myr,
corresponding to a plausible estimate of the oldest
stars in the cluster (see Barrado y Navascu\'es et al
2005 and references therein).
We  fitted both black bodies and theoretical atmospheric models 
(Allard et al. 2001, 2003) to the SEDs 
and derived  T$_{eff}$=2125 K, in full
agreement with the values predicted by the photometry, the isochrones, and the 
spectral type. Thus, we are confident of the membership of
 this object and its substellar nature.

\begin{figure*}
\includegraphics[angle=0,scale=.60]{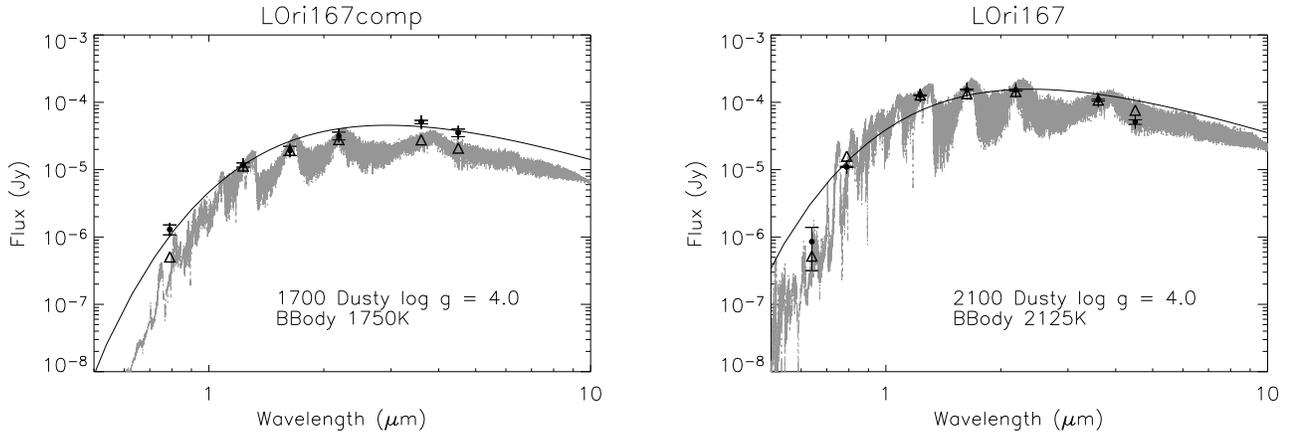}
\caption{Spectral energy distributions for our targets, 
using optical, near-IR and Spitzer/IRAC data.  Solid circles represent the actual measurements, 
and open triangles correspond to the synthetic photometry from the models.}
\end{figure*}


The visual companion (LOri167comp) is
 probably too faint for useful low-resolution IR spectroscopy ($J$=20.19).
If a member, it would probably be an object with mid to late L spectral type.
An SED fitting indicates T$_{eff}$=1750 K, corresponding to L6 
on the temperature scale of Basri et al. (2000). 
This temperature and the position in the color-magnitude 
diagrams fully agree with membership.
The predicted mass, using  a 5 Myr Dusty  isochrone by the Lyon group,
 is in the range 7--13 $M_{jup}$, with an optimal value, as derived from  T$_{eff}$,
 of 8 $M_{jup}$. A 10 Myr isochrone (much older than the age of the cluster) 
would produce masses in the range 11--15 $M_{jup}$.
%
The high values in mass come from the Spitzer/IRAC magnitudes and the Dusty  isochrone
might indicate the possible presence of a disk or ring, although the
 magnitudes in the theoretical models 
are far from perfect, especially in the almost unexplored 
domain of the extremely faint,  red, and low mass objects.

\begin{table*}
\small
\caption[]{Data for the LOr167 visual binary. }
\begin{tabular}{lllllllll}
\hline
LOri    & alpha (2000.0) delta   & R$^{(1)}$  &    I$^{(1)}$&  J$^{(2,3)}$& H$^{(2)}$  & Ks$^{(2)}$ & [3.6]      & [4.5]   \\
\hline				    				    				    
167     & 83.8091354   9.9020901 & 23.86 0.64 &  20.90 0.02 &  17.88 0.03 & 17.15 0.03 & 16.62 0.03 & 16.04 0.03 & 16.37 0.09  \\
167comp & 83.8104782   9.9020519 & --         &  23.23 0.18 &  20.19 0.19 & 19.41 0.17 & 18.31 0.14 & 16.85 0.06 & 16.76 0.14  \\
\hline
\end{tabular}
$\,$\\
$^{(1)}$ $RI$ photometry from CFHT September 1999 (Barrado y Navascu\'es et al. 2004).\\
$^{(2)}$ $JHKs$ data from WHT/INGRID November 2002.\\
$^{(3)}$ $J$=17.78$\pm$0.02 and 20.36$\pm$0.08 for the LOri167  and LOri167comp  in the deep J image (WHT/INGRID February 2003). \\
LOri167 J=18.01$\pm$0.03, H=17.17$\pm$0.07, Ks=16.83$\pm$0.09,  [3.6]=15.935$\pm$0.063,  [4.5]=16.060$\pm$0.129,
in Barrado y Navascu\'es et al. (2007). The near IR data were taken with Omega2000 at Calar Alto Observatory (October 2005).
\end{table*}

\subsection{Are the two objects a physical pair?}

We  estimated the contamination by unrelated visual companions by
deriving the probability of finding an object with the luminosity and colors of LOri167comp.
The total number of objects brighter by half a magnitude in both I and J, and with I-J$>$2.8,
 is equal to three.
This estimate is very conservative since the maximum uncertainties in each band are 
below  0.16 mag in $I$ and 0.07 mag in $J$. Counting only the objects
with $I$ and $J$ magnitudes and $(I-J)$ colors within these uncertainties, we find only one object, 
which is LOri167comp itself. 
Since our total field of view is  4.1\arcmin$\times$4.1\arcmin, 
the probability of finding such  a faint and red object in a 5 arcsec circle around any 
object is three times 1.3$\times$10$^{-3}$, about 0.4\%.
In this field of view, there are 11 probable members of the Collinder 69 cluster
(Barrado y Navascu\'es, Hu\'elamo, Morales-Calder\'on 2005; Barrado y Navascu\'es et al. 2007),
 but only three are in the substellar domain (such as LOri167), with magnitudes less than $Ic$=17.55. Therefore,
the total probability of finding a visual binary by chance  mimicking 
this type of binary (two brown dwarfs or a brown dwarf
 plus a planetary-mass object) is slightly higher than  1\%.
Because this probability is reasonably low, we believe
it favors the planetary-mass object  being a real companion.  
 Such a calculation is only tentative and must be considered with caution.
We are  well aware of the dangers of applying
a posteriori statistical tests, so are simply arguing that
the data are consistent with the planetary-mass candidate being a companion,
although it will be difficult to prove (or disprove) this
association with existing instrumentation.
And we emphasize that LOri167comp
is well below the completeness limit of our optical and infrared surveys.

Therefore,  this probabilistic argument based on  photometric analysis of the objects in the field
supports the hypothesis that
it is, indeed, a physical pair formed by a low mass brown dwarf and a 
planetary-mass object. 
A population of low-mas binaries with a 100-250 AU separation has been discovered
in UpSco and other young associations (Bouy et al. 2006 and references therein)
or in the field (Bill{\`e}res et al. 2005). 
Moreover, Caballero et al. (2007) have
found  that the visual binary SE70$+$SOri68 might be real, with a separation of 4.7\arcsec,
equivalent to 1700 AU.  Due to the separation and low mass, even if they are bound,
most of  these wide binaries  will be disrupted soon, either by the interactions with other cluster members or
by field stars and the host association moving in the galactic gravitational field.
However, Caballero (2007) has recently discovered two low mass stars with common proper motion and a projected
separation of  1800 AU.
During the revision of this paper, the widest low mass binary was announced, with a projected 
separation  of about 5,100 AU and a mass for the primary lower
 than 0.1 M$_\odot$ (Artigau et al. 2007).
Since this seems to be an old, field system with stellar masses, 
it suggests that at least a few wide binaries may survive
for very long periods of time. 
Additional results have also been published by Kraus \& Hillenbrand, (2007) in three young associations.
In any case, LOri167 would have a binding energy about 20 times lower than Oph\#11 
(Jayawardhana \& Ivanov 2006; Close et al. 2006), 
making it unlikely that the LOri167
binary would survive to an old age.

The very existence of this type of  pair is very important, because it would
prove that both very low mass brown dwarfs and planetary-mass objects form 
from fragmentation and collapse, even if the process takes place in a hierarchical
process, with successive fragmentations generating 
binary or multiple systems.
The embryo hypothesis cannot be completely ruled out as a complementary mechanism
to form very low mass brown dwarfs and planetary-mass objects,
 but due to the evolutionary stage of 
LOri167 and other similar systems, it could not explain
these types of binaries at their wide separation.


\section{Conclusions}

By collecting multi-wavelength photometry in the range 0.8-4.5 micron, we
have identified a very low mass visual binary in the 5 Myr Lambda Orionis 
cluster, separated by 5\arcsec and equivalent to 2000 AU.
We have confirmed the brown dwarf nature of the primary, 
whose mass is in the range 15-20 $M_{jup}$ and
T$_{eff}$=2125 K.
The secondary, with T$_{eff}$=1750K,  would have a mass in the range 
 7-13 $M_{jup}$ if it were a  member of the cluster.
This system is  very important since it imposes strong constraints
 on the proposed formation scenarios for brown dwarfs,
such as turbulent fragmentation (Padoan \& Nordlund 2004), 
 ejection from multiple proto-stellar systems 
(Reipurth \& Clarke 2001 or Bate et al. 2002), or
a photoevaporation of  massive pre-stellar cores 
(Whitworth by \& Zinnecker 2004).

\begin{acknowledgements}
This research has been funded by Spanish grants  MEC/ESP2004-01049, MEC/Consolider-CSD2006-0070, 
and CAM/PRICIT-S-0505/ESP/0361.
\end{acknowledgements}



\begin{thebibliography}{}
%
%
\bibitem[Allard et al.(2001)]{2001ApJ...556..357A} Allard, F., Hauschildt, 
P.~H., Alexander, D.~R., Tamanai, A., \& Schweitzer, A.\ 2001, ApJ, 556, 
357 
\bibitem[]{}
Allard N.F., Allard F., Hauschildt P.H., Keilkopf J.F:, Machin L.,
2003, A\&A Letters 411, 473
%
\bibitem[]{}
Artigau E.,  Lafreni\`re D,  Doyon R.,  Albert L,  Nadeau D.,  Robert J.,
2007, ApJ Letters, in press (astro-ph/0702647)

\bibitem[]{}
 Baraffe, I., Chabrier, G., Allard, F., Hauschildt, P.H.:  1998, A\&A 337, 403
%
\bibitem[]{}
 Baraffe, I., Chabrier, G., Allard, F., Hauschildt, P.H.: 2002, A\&A 382, 563
%
\bibitem[]{} 
Barrado y Navascu\'es, D., Stauffer, J.R., Bouvier, et al.
2004, ApJ 610, 1064
%
\bibitem[Barrado Y Navascu{\'e}s et al.(2005)]{2005imf..conf..133B} 
Barrado  y Navascu{\'e}s, D., Stauffer, J.~R., \& Bouvier, J.\ 2005, ASSL Vol.~327: 
The Initial Mass Function 50 Years Later, 133 
%
\bibitem[]{} 
Barrado y Navascu\'es D. Hu\'elamo N, Morales-Calder\'on M. 2005, AN 326, 981
%
\bibitem[]{}
Barrado y Navascu\'es D., Stauffer J.R., Morales-Calder\'on M., et al.
2007, ApJ, in press
%
\bibitem[Basri et al.(2000)]{2000ApJ...538..363B} 
Basri, G., Mohanty, S., 
Allard, F., Hauschildt, P.~H., Delfosse, X., Mart{\'{\i}}n, E.~L., 
Forveille, T., \& Goldman, B.\ 2000, \apj, 538, 363 
%
\bibitem[2002]{Bate2002}   
Bate, M.R., Bonnell, I.A., Bromm, V.,  
2002, MNRAS 332, L62 
%
\bibitem[Bill{\`e}res et al.(2005)]{2005A&A...440L..55B} 
Bill{\`e}res, M.,  Delfosse, X., Beuzit, J.-L., Forveille, T., Marchal, L., \& Mart{\'{\i}}n, E.~L.\ 
2005,  A\&A Letters, 440, L55 
%
\bibitem[]{} 
Bouy  H,  Martin E.L., Brandner W. et al.
2006, A\&A 451, 177
%
\bibitem[]{}
Caballero J.A:, Martin E.L., Dobbie P.D., Barrado y Navascu\'es D.,
2007, A\&A in press (astro-ph/0608659)
%
\bibitem[]{}
Caballero J.A.,
2007, A\&A Letters 462, 61
%
\bibitem[]{}
Chabrier G., Baraffe I., G., Allard, F., Hauschildt, P.H.: 2000, ApJ 542, 464
%
\bibitem[]{} 
Chauvin G, et al.,  2004, A\&A Letters 425, 29
%
\bibitem[]{} 
Close L. M., Siegler N., Freed M., Biller B., 2003
ApJ 587, 407
%
\bibitem[]{} 
Close et al. 2006, ApJ, in press (astro-ph/0608574) 
%
\bibitem[{{Dolan} \& {Mathieu}(1999)}]{Dolan99}
{Dolan}, C.~J. \& {Mathieu}, R.~D. 1999, \aj, 118, 2409
%
\bibitem[{{Dolan} \& {Mathieu}(2001)}]{Dolan01}
{Dolan}, C.~J. \& {Mathieu}, R.~D. 2001, \aj, 121, 2124
%
\bibitem[]{} 
Jayawardhana R., \& Ivanov V., 2006, Science 313, 1279

\bibitem[Kraus et al.(2005)]{2005ApJ...633..452K} 
Kraus, A.~L., White,  R.~J., \& Hillenbrand, L.~A.\ 
2005, \apj, 633, 452 

\bibitem[Kraus et al.(2006)]{2006ApJ...649..306K} 
Kraus, A.~L., White,  R.~J., \& Hillenbrand, L.~A.\ 
2006, \apj, 649, 306 

\bibitem[]{} 
Kraus, A.L.,   \& Hillenbrand, L.~A.\
2007, ApJ, in press (astro-ph/0702545)
%
\bibitem[]{} 
Luhman K.L.,  2004, ApJ 614, 398
%
\bibitem[]{} 
Luhman K.L.,  2005, ApJ Letters 633, 41
%
\bibitem[]{} 
 Padoan P. \& Nordlund 2004, ApJ 617, 559, 
%
\bibitem[2001]{reipurth2001} 
Reipurth, B., \& Clarke, C., 
2001, AJ 122, 432 
%
%
\bibitem[]{} 
Whitworth by \& Zinnecker 2004, A\&A 427, 299
%
\bibitem[]{} 
%
\end{thebibliography}
\end{document}